# Making Sense of Physics through Stories: High School Students' Narratives about Electric Charges and Interactions

Ying Cao

Abstract: Educational research has shown that narratives are useful tools that can help young students make sense of scientific phenomena. Based on previous research, I argue that narratives can also become tools for high school students to make sense of concepts such as the electric field. In this paper I examine high school students' visual and oral narratives in which they describe the interaction among electric charges as if they were characters of a cartoon series. The study investigates: given the prompt to produce narratives for electrostatic phenomena during a classroom activity prior to receiving formal instruction, (1) what ideas of electrostatics do students attend to in their narratives?; (2) what role do students' narratives play in their understanding of electrostatics? The participants were a group of high school students engaged in an open-ended classroom activity prior to receiving formal instruction about electrostatics. During the activity, the group was asked to draw comic strips for electric charges. In addition to individual work, students shared their work within small groups as well as with the whole group. Post activity, six students from a small group were interviewed individually about their work. In this paper I present two cases in which students produced narratives to express their ideas about electrostatics in different ways. In each case, I present student work for the comic strip activity (visual narratives), their oral descriptions of their work (oral narratives) during the interview and/or to their peers during class, and the their ideas of the electric interactions expressed through their narratives.

The concept of the electric field is emphasized in high school and college physics curricula, but students often have difficulties learning it. Past research in physics education has examined high school and college students' understandings of the concept and/or theory of electric field by adopting survey assessments. The literature has reported students' learning difficulties based on the results of these assessment. Noticing that most of the assessment items adopted in past research were written using canonical terms, symbols, and diagrams, some researchers have contemplated that the formal representations used in the assessments might have constrained students expressing their ideas (e.g., Maloney, O'Kuma, Hieggelke, & Van Heuvelen, 2001). As Maloney and colleagues admitted, many of the questions in their assessments were written using formal physics terms, which made it difficult to know how students actually interpreted the questions on the tests. Maloney and colleagues pointed out that students might not interpret the vocabulary in the same way as the designers of the questions did. Therefore, Maloney and colleagues speculated that students' seemingly wrong ideas perhaps stemmed from the use of vocabulary and phrases they did not really understand.

The present study attempts to remove the constraints of formal language and asks students to express their ideas of the electric field in plain, everyday language. Informed by the educational research literature in which scholars have found that narratives can empower elementary and middle school students making sense of scientific concepts that are often difficult to learn (e.g., Ochs, Taylor, Rudolph, & Smith, 1992; Sohmer & Michaels, 2005), I decided to look at high school students' narratives and the ideas of the electric field that they express through these narratives.

In the present study, a narrative refers to a story (Sohmer and Michaels, 2005). A narrative includes the following elements: characters (who), time (when), location (where), and a series of events (what happened). In the present study, students were asked to produce a narrative about electric charges (as characters) and electric interactions (as events). They were asked to describe the narratives about electric charges in two formats (Keats, 2009): oral narratives and "visual narratives" (Cohn, 2013, p. 8). Here, I argue that oral and visual narratives can become tools for high school students to make sense of concepts such as the electric field. In this paper I explore the questions: given the prompt to produce narratives for electrostatic phenomena during a classroom activity prior to receiving formal instruction,

(1) What ideas of electrostatics do students attend to in their narratives?

(2) What role do students' narratives play in their understanding of electrostatics?

## Background

**Narratives in Science Education**

Scholars in science education who work with young students (such as in elementary and/or middle school) view narratives as useful learning tools (e.g., Ochs, Taylor, Rudolph, & Smith, 1992; Sohmer & Michaels, 2005). Sohmer and Michaels (2005) investigated middle school students' use of narrative tools to explain scientific phenomena. Sohmer and Michaels (2005) argued that narrative tools play a far more important role in science education than the literature would suggest. In their work, Sohmer and Michael studied sixth-grade students who learned a specific, teacher-introduced narrative tool—the "Two Puppies" story—in which "puppies" are used as a metaphor for air molecules. The authors drew the notion of narrative or story from Bruner's account of narrative mode (in contrast with logico-scientific or diagrammatic mode) that involves "characters in action with intentions or goals in settings using particular means" (Bruner, 1986, p. 20). They drew from Bruner's definition the elements of character and action. As they explained,

> In the "Two Puppies" story, the "puppies" referred to are mythical or fictional beings—"Air-puppies"—combining *some* of the properties of real, live puppies with the behavioral characteristics of air molecules. The air-puppies are the bumbling (mindless) agents in a modifiable story with a particular setting (always including two rooms separated by a moveable wall-on-wheels), participating in a series of events, always resulting in some kind of lawful effect—that is, the wall moves as it *must*, given the air-puppies opposing impacts upon both sides. (p. 63)

Sohmer and Michaels' study results showed that students successfully learned to use the "The Two Puppies" story as a tool to explain phenomena of air pressure in the kinetic theory of gases.

The learning activity featured in the present study is a comic strip drawing activity that I will describe in more detail below. The comic strip activity has similarities and differences with Sohmer and Michaels' work (2005). In the comic strip activity electric charges are described in the worksheet as animated characters (similar to air molecules in Sohmer and Michaels' work). Electric charges are put into a setting: a room with a door to let a test charge come in (similar to Sohmer and Michaels' rooms with a movable wall for air molecules to hit and push). Electric charges take part in a series of events that demonstrate physical laws, such as like charges repel and opposite charges attract (similar to Sohmer and Michaels' work in which air molecules take part in events and demonstrate the kinetic theory of gases). However, in contrast to Sohmer and Michaels' work, in the present study students are given the opportunity to choose particular metaphors with particular features as their individual tools to describe electric charges and interactions (different from the "two puppies" tool that was taught to the students in Sohmer and Michaels' study). I will elaborate more on the comic strip activity in later sections of this paper.

In their work about the role of narratives in science education, Ochs, Taylor, Rudolph, and Smith (1992) focused on storytelling, which they describe as "one type of narrative activity" (p. 69). The authors pointed out the important role of collaborative storytelling in theory building. According to Ochs and colleagues complex storytelling in which perspectives are challenged and redrafted collectively is more likely to occur when co-narrators are familiar with one another and/or with the narrative events. In their study Ochs and colleagues concluded that for children in elementary and secondary school sitting together, listening to each other, and collaborating on a certain topic, theory-building can be fostered through this kind of narrative activity (Ochs, Taylor, Rudolph, & Smith, 1992).

During the comic strip activity in the present study, a group of high school students produced their individual narratives of electric charges and shared the narratives with their peers. Students' opinions on their peers' work (e.g., whether a story makes sense) expressed during the activity also may lead them to think and talk more about the electric field, just like in the study carried out by Ochs and colleagues (1992).

**Narrative Explanations**

Studies have also examined narrative explanations in written professional scientific texts (e.g., Norris, Guilbert, Smith, Hakimelahi, & Phillips, 2005). Narrative explanations in these studies refer to the narratives that explain scientific phenomena and/or theories. Norris and colleagues drew on Herrnstein Smith's definition of narrative discourse—verbal acts consisting of "someone telling someone that something happened" (1981, p. 228)—and highlighted from that definition the following elements: a narrator (someone telling), a narratee (someone receiving), events (something that happened), and past time (something happened in the past). In addition to these elements, Norris and colleagues pointed out other important narrative features such as narrative appetite, structure, agency, and purpose. Based on a review of previous work, Norris and colleagues developed a framework about narrative explanations in which they highlighted the relationship among/between *events.* According to Norris and colleagues, a narrative explanation:

- Explains an event by narrating the events leading up to its occurrence;
- Cites unique events as explanatory of other unique events;
- Posits some events as causes of others;
- Seeks unification (but does not supply deductive tightness) by showing how the event to be explained is one of an intelligible series of events.

(Norris et al., 2005, p. 550)

Norris and colleagues argued for the supposedly desirable effects of using narrative explanations in science education: to improve memory for content, to enhance interest in learning, and to enlarge comprehension of what is learned. Although the authors pointed out these benefits, they nonetheless included in their list of features of a narrative explanation what it does not do: it

Rarely supports predictions, but rather relies upon retrodiction to indicate how the present is a consequence of the past. (Norris, et al., 2005, p. 550).

Norris and colleagues' definition of a narrative explanation reflects their position that narrative explanations play a positive, but limited, role in explaining scientific phenomena.

In the present study, I focus on examining narratives students produce in the comic strip activity. The narratives students produce do not have to explanations but could be descriptions about the interaction between electric charges.

**Visual Narratives**

In linguistics, Cohn, in his book about visual language of comics, calls comic strips "visual narratives" (Cohn, 2013, p. 8). Keats, from a discourse analysis perspective, also argued that visuals (e.g., images, photos) are important analytical foci to narrative research. Keats developed a methodology about "multiple texts analysis in narrative research" (2009, p. 181). Keats addressed "three types of narrative texts," which includes "visual, written, and spoken stories" (p. 181). Keats argued that in research about narratives, imagery productions "are very helpful and useful when participants have difficulty recording emotions, impressions, or aspects that were difficult to put into words" and "can also expand a researcher's opportunity to better understand the complex narrative participants construct" (Keats, 2009, p. 187).

Visual narratives, and/or multiple types of narratives including visual texts, have not been well studied in educational research. The present study looks at the oral and visual narratives that students produce about electric charges. Students' drawings during the comic strips activity are their visual narratives. Students' spoken words when they describe their comic strips are their oral narratives. The inclusion of both oral and visual narratives can allow students to recruit more modalities to express their ideas. It also expands the researchers' opportunities to learn about students' ideas compared to only looking at student work in verbal formats (such as classroom speech and/or work in words).

**My Study**

The present study examines high school students' visual and oral narratives in which they describe the interaction among electric charges as if they were characters of a cartoon series. Compared to previous studies, the present study extends the literature in the following ways. First, the participants (high school students) bridge the age gap in previous studies on use of narratives that either focused on elementary and middle school students or on adult scientists. Generally, narratives are less promoted in classrooms and less studied by educational researchers among older, high school students. Second, I focus on student-produced narratives prior to formal instruction (Sohmer and Michael's study focuses on students' use of the teacher-introduced "little puppies" story after instruction). Third, my study examines data of multiple narrative formats (visual and verbal narratives).

I chose to focus on the subject of electrostatics because the electric field and the concepts related to it have been deemed too abstract and hard to learn (e.g., Furio & Guisasola, 1998; Maloney, O'Kuma, Hieggelke, & Van Heuvelen, 2001; Rozier & Viennot, 1991). A fair amount of research has shown that students have difficulties understanding the theories (Maloney, O'Kuma, Hieggelke, & Van Heuvelen, 2001; Rozier & Viennot, 1991), concepts (Furio & Guisasola, 1998), and/or representations (Tornkvist, Pettersson, & Transtromer, 1993) related to the electric field. This motivated me to tackle the subject using a different approach: producing narratives. Trying to produce narratives about electric charges, students are expected to engage in making sense of the phenomena. The narratives that students construct might serve as a tool for students to express their understandings, for educators to access students' conceptions and help students learn, and for educational researchers to understand how they learn.

## Theoretical Framework

This work is grounded on the overlapping area of a few inter-related frameworks in educational research. My interests in learning about student thinking prior to instruction are rooted in the constructivist tradition of educational research. I follow the theoretical strands that claim that students' ideas are idiosyncratic and piecemeal (diSessa, 1988) and that they constitute precious learning resources (Hammer, Elby, Scherr, & Redish, 2005). Based on these theories, learning is building conceptions on the resources learners possess, and it is informative and important for educators to access students' learning resources before taking instructional actions.

One way to access students' ideas and/or learning resources is to examine the discourse students construct while they participate in a learning activity. According to scholars of discourse analysis, language-use reflects the ways people think, act, and interact with other people in a certain context (e.g., Gee, 2005). I adopt this perspective and apply it to an educational setting (a high school physics classroom) and examine a specific kind of discourse (narratives). In the present study, students produce narratives (in words and drawings) of electric charges and interactions. Examining the narratives in detail can allow me to learn about students' ideas of electric charges and interactions in a way that traditional paper and pencil assessments might not afford.

Based on these theories I formulate two research questions for the present study: given the prompt to produce narratives for electrostatic phenomena during a classroom activity prior to receiving formal instruction,

(1) what ideas of electrostatics do students attend to in their narratives?

(2) what role do students' narratives play in their understanding of electrostatics?

Exploring these questions I aimed to find the connections between the narratives students produced and the ideas of the electric field students expressed to support my argument in this paper: oral and visual narratives can become tools for high school students to make sense of concepts such as the electric field.

**Methods**

**Participants**

Participants in my study were a group of 36 high school students enrolled in a summer program in China. At the time they participated in the study, they were between 15 and 16 years old, had completed the ninth grade, and were taking summer courses to prepare for the tenth grade (which is the first year of high school in China). They had not yet received formal instruction on electrostatics. I was one of the physics teachers in the summer program.

**The Comic Strip Activity**

The physics course I taught consisted of 12 sessions lasting three weeks. During the 12-session course, I covered a unit of algebra-based one-dimensional motion, a unit about the concept of force and related vector composition, and a unit of electricity. The electricity unit was carried out in the last five sessions (during the last five days) of the course in parallel with the mechanics content on the same days. The electricity lessons consisted of two learning activities: the Electric Field Hockey activity[1] (for the first three days) and the comic strip activity (for the remaining two days). The goal of the Electric Field Hockey activity was to expose students to electrostatic phenomena and prepare them for the comic strip activity, which is the focus of this paper.

During the comic strip activity, students were asked to draw comic strips that showed interactions between positive and negative charges as if they were characters in a cartoon series (the initial scene of each story was described on the worksheet; see Table 1). I designed and implemented the comic strip activity for a number of reasons. First, the task directly prompts students to construct narratives (described as "stories" in the task requirement) of electric charges and share the narratives with their peers. Second, the inclusion of visual narratives can allow students to recruit more modalities to express their ideas, and can expand the researchers' opportunities to learn about students' ideas compared to only looking at student work in verbal formats (such as classroom speech and/or work in words). Third, the worksheet served as a way to keep track of students' understandings in a written form that would augment the videotaped recording of the ideas students expressed in class. Fourth, since high school students are generally interested in reading comic books, I believed that asking them to draw comic strips might increase their enthusiasm towards the activity. Finally, because students are asked to draw comic strips, the task placed no vocabulary constraints on them; therefore, I stood a better chance at capturing expressions of their thinking regarding electrostatic phenomena.

---

[1] During this activity students played a Web-based game called Electric Field Hockey downloaded from the website produced by PhET, a University of Colorado educational technology group that designs science simulations for classroom use (http://phet.colorado.edu/en/simulation/electric-hockey). An analysis about students' participation in the E-Hockey activity is the focus of Chapter 5 of this dissertation.

In Sections 1, 2-1, and 2-2, the tasks were asking students to draw four frame comic strips. The tasks set up students to draw multiple frames for each story thus could cue students to describe the process of the electric interactions, not just to describe a single state. Four frame comic strips is a common structure of comic stories. High school students are likely to already be familiar with this structure and should feel easy to get started with these tasks. In Sections 3-5, the tasks were no longer asking students to draw comic strips but to show the impact of electric charges on a test charge in one single frame. The tasks in Sections 3-5 aim to probe students' thinking about electric fields (rather than electric forces that were probed in Sections 1, 2-1 and 2-2). Table 1 summarizes the tasks in the comic strip activity worksheet.

Table 1. Comic Strip Activity Tasks

| Text | First Scene |
|------|-------------|
| Section 1<br>Draw Something: Draw a four-frame comic telling a story about two characters: Little Positive (P) and Little Negative (N). P is sitting at the center of a room when N comes in through the door. What will happen next? Use four frames to draw at least four scenes of the story. The picture shown could be the first panel, but feel free to draw a picture for the first panel. You are not required to draw every character vividly; the key is to show how and why your characters act the way they do. | 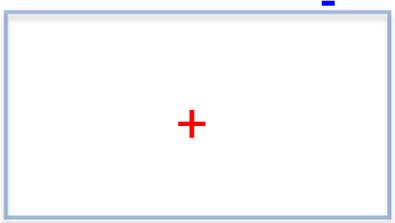 |
| Section 2-1<br>Draw Something: Draw a four-frame comic telling the stories of two big charges and one little charge. For this first story, there is one Big P, one Big N, and one Little P. Big P and Big N are sitting in the room at a distance. They are fastened to their seats and can't move. Little P comes into the room. What will happen next? Explain your story to the class. What do all the things you have drawn mean? | 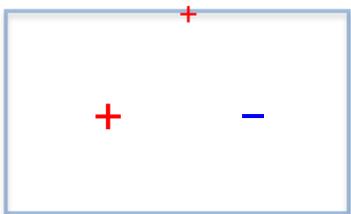 |
| Section 2-2<br>Draw Something: Draw a four-frame comic telling stories of two big charges and one little charge. For this second story, both big charges are negative, while the little charge is positive. What will happen when the little charge enters the room? Explain your story to the class. | 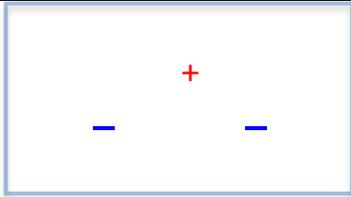 |
| Section 3<br>Draw Something: Now what if we take away the Little Positive charge from the previous two pictures? Can you find a way to show the influence they would have on the little positive charge that were present in Section 2-1 and 2-2? | 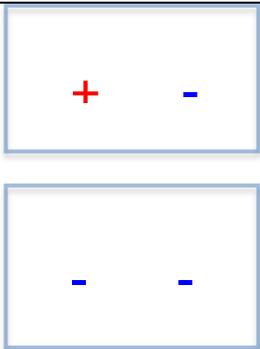 |
| Section 4<br>Draw Something: What if we leave only one big charge in the room, P or N? How can you show the influence the big charges (Big Positive or Big Negative) would have, respectively, on a Little Positive charge that might come into the room? | 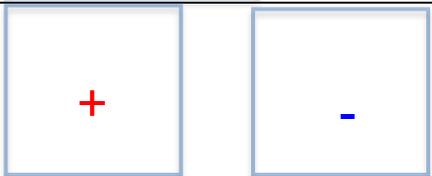 |

| Section 5<br>Based on what you have drawn before, how can you show the influence the two charged plates would have on a Little Positive charge that moves between the plates? Why is this true? | 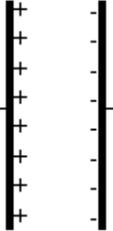 |

There were five consecutive sections in the comic strip activity. Sections 1, 2-1, and 2-2 asked students to draw a four-frame story for each starting scenario that was provided. In Section 1, a big charge sits in the room and is immobile. Then a little negative charge comes into the room. The students were asked to draw a story about what happened next. In Section 2-1, a big positive charge and big negative charge sit in the room and are unable to move. And in Section 2-2, two big negative charges sit in the room. In the scenarios for both Sections 2-1 and 2-2, a little positive charge comes into the room. Again, students were asked to draw what happened next. Sections 3 and 4 asked students to show the influence the big charges would have if a little positive charge were present. Section 3 used the two scenarios of Sections 2-1 and 2-2, and Section 4 put the big positive charge and the big negative charge in separate rooms. Section 5 asked students to show the influence of charged parallel plates on a test charge.

**Lesson Structure**

At the beginning of the first day of this two-day activity, I handed out packets of worksheets to each individual student. I explained to the students that this activity was open-ended, and there were no right or wrong answers. I also told students that this activity was part of my research project.

Overall, students spent 80 minutes on this activity (50 minutes on the first day and 30 on the second day). Table 2 shows the activity format and the timeline. The time codes are continuous from day one to day two.

Table 2. Activity format and timeline.

| Day | Section | Format | Start and stop times | Duration |
|---|---|---|---|---|
| Day 1 | Section 1 | Individual work | 00:00:00–00:12:30 | 12:30 minutes |
| | | Small-group sharing | 00:12:30–00:17:14 | 4: 44 minutes |
| | Sections 2-1 and 2-2 | Individual work | 00:17:14–00:30:16 | 13:02 minutes |
| | | Small-group sharing | 00:30:16–00:34:00 | 3: 44 minutes |
| | | Whole-group sharing | 00:34:00–00:42:50 | 8:50 minutes |
| | Section 3 | Individual work | 00:42:50–00:50:00 | 7:10 minutes |
| Day 2 | Section 4 | Individual work | 00:50:00–00:58:00 | 8:00 minutes |
| | | Small-group sharing | 00:58:00–01:01:04 | 3:30 minutes |
| | | Whole-group sharing | 01:01:04–01:11:26 | 10: 22 minutes |
| | Section 5 | Individual work | 01:11:26–01:15:50 | 4: 24 minutes |
| | | Small-sharing | 01:15:50–01:20:00 | 4: 10 minutes |

As shown in Table 2, the activity was structured to accommodate individual work, small-group sharing, and whole-group sharing. First, students worked individually on Section 1, and I walked around answering students' questions. After most of them had finished, I asked students to form small-groups with neighbors and share their work. Then students were asked to work on Sections 2-1 and 2-2 (treating it as one single section) individually and then share their work with a small group. After they discussed their work with their group for a few minutes, I randomly picked a line of six students and projected their work on the board one by one. The whole class read the work of each of these six students on the board while the student described their comic strips to the class. After that, the class moved on to work on the next section.

At minute 50 into the activity, I announced the end of the session, collected student work, and dismissed the class. On the following day, I handed back the students' work, and they continued to work on Sections 3, 4, and 5 for 30 minutes. The format was similar to that of the first day. By the end of the day, most of the students had completed all sections. When I started to collect the work at the end of the second day and then dismissed the class, most students were in the middle of small-group discussions of Section 5.

**Data Collection**

　　Data for this study were collected from three sources. The first source was the written student work for the comic strip activity. The second source was the video documentation of the class as a whole and of a randomly-selected small group: Zixuan, Mengdi, Hanzhong, Ru, Yatong, and Yifan. The class was seated in a traditional matrix (rows and lines) facing the blackboard. Each student in the class had been randomly assigned to a fixed seat at the beginning of the summer session. The six students in the focus group were seated one next to another in two rows and three lines. I picked this group to videotape because they were seated on the side and front of the classroom and there was more room to set up multiple cameras around them. They and their parents had all consented to be videotaped for my study. No other criteria were considered for choosing this group as the focus of the present study. The third source came from one-on-one post-class interviews I carried out with these six students. I videotaped the whole class with two cameras; one at the front of the room focused on the students and one at the back of the room focused on myself, the teacher. In addition, the randomly selected group of six students was recorded with three cameras around them throughout the comic strip activity.

　　I collected 36 student worksheets after the activity was completed. During the post-class interviews, the six selected students provided me with in-depth explanations of their drawings and/or writing on the worksheets. The interviews were recorded by two cameras: one focused on the student's face and one on his or her hands and work on the desk. I carried out the interviews in Chinese. Each interview consisted of two parts. Part one was the student explaining his or her own work to me section by section. During part two the interviewed student was asked to read and comment on the other five students' work. The six students were interviewed on different dates. I first interviewed Mengdi. During our interview, I first asked her to describe her own work of Section 1 and then to comment on other students' work of Section 1. Then we moved on to the next section, and so forth. Therefore, the two parts of the interview were not fully separated in Mengdi's interview (but were separated within each section). I realized that this sequence might make it hard for me to separate the student's own ideas from the new ideas she developed by reading and commenting on other students' work during the interview. Hence, in the rest of the interviews, I first asked the student to describe his or her own work on all sections and then asked him or her to read and comment on other students' work. Yatong was interviewed on the day after Mengdi was interviewed. Zixuan, Yifan, and Ru were individually interviewed on the day after Yatong was interviewed. Hanzhong was interviewed several days later after the summer program had ended. The analysis for this paper mainly focuses on part one of the interview. Data from part two of the interview are used as supplementary data if they provide more information and help me explain students' ideas.

The classroom videos, student written work, and post-class interview videos provided the data for my analysis. These data allowed me to explore the narratives students produced about electric charges and their understandings of electrostatics.

**Data Analyses**

In the present study, I examine students' narratives (spoken words and drawings) and identify narrative elements and physical ideas mentioned in the narratives. For this purpose, I developed six case studies with the students in the small group based on the following data set: (1) student work during the comic strip activity; (2) classroom videos recording the whole group and the small group participating in the comic strip activity; and (3) interview videos with each of the six students in the small group. To start my analysis, I created electronic versions of students' written work and transcribed the interviews and classroom videos verbatim.

I started off with a line-by-line examination of the interview and classroom transcripts, together with reading and annotating on the student work that students were referring to in the transcript. I took notes about the students' description (to me and to their peers) of each narrative they produced in their work. During this process, I tried to understand the general narrative they had constructed for the electric charges.

I then proceeded with a second round of reading of the interview transcripts, identifying the elements in each narrative: characters and relationships (who), time (when), location (where), and events (what happened). The source I focused on to identify these elements was their drawings of the comic strips and the transcripts where they orally described the narrative to me and to their peers.

After that, I examined the data a third time to identify their explicit reference to the physical ideas about electrostatics. During this round, I focused on identifying the law-like statements students brought up about physical concepts and/or theories. For example, I tagged places in the transcripts where students mentioned attraction, repulsion, force, distance, strength, and/or statements such as "like charges repel, and opposite charges attract." I identified these descriptions and marked the place students brought them up in relation to specific narrative elements.

In sum, I tagged in the transcript the narrative elements: characters and relationships (who), time (when), location (where), and events (what happened). I also tagged the physical ideas stated in the same transcript. For example, in the following transcript:

*Because like charges repel, the dad and the son are both positive charges, they stand far away from and are mad at each other.*

The phrase *like charges repel* was marked as a physical idea, *dad and son* was marked as character relationship in the story, and *stand far away, mad at each other* were marked as events.

During each round of analysis described above, I examined all six cases one by one. At the end of each round of analysis, I checked across the six cases to see if there was any additional information for me to understand a particular case. For example, when students were describing their work to me during the interview, they often mentioned some moments in class when they argued about a particular student's work (e.g., whether a part of the drawing made sense to them and why). I checked across all the six students' data to see if they described the moment the same way. After all three rounds of analysis were complete, I constructed a comprehensive, thick description for each case. Table 3 summarizes the process of data analysis.

Table 3. Data analysis.

| Round 1: understanding narratives | | Round 2: identifying narrative elements | | Round 3: identifying physical ideas | | Wrapping-up |
|---|---|---|---|---|---|---|
| Case 1 | Cross-case checking | Case 1 | Cross-case checking | Case 1 | Cross-case checking | Comprehensive description of case 1 |
| Case 2 | | Case 2 | | Case 2 | | Comprehensive description of case 2 |
| Case 3 | | Case 3 | | Case 3 | | Comprehensive description of case 3 |
| Case 4 | | Case 4 | | Case 4 | | Comprehensive description of case 4 |
| Case 5 | | Case 5 | | Case 5 | | Comprehensive description of case 5 |
| Case 6 | | Case 6 | | Case 6 | | Comprehensive description of case 6 |

**Data Selection**

In this paper I present two cases: Zixuan's and Mengdi's cases. Zixuan and Mengdi produced narratives in different ways. Zixuan was one of the three students (Zixuan, Yatong, and Yifan) in this small group who drew visual narratives when the task was to draw comic strips (Sections 1, 2-1, and 2-2) and shifted to drawing diagrams when the task did not explicitly ask to draw comic strips (Sections 3, 4, and 5). For this reason, I will only describe the visual narratives in Zixuan's work for Sections 1, 2-1, and 2-2 in the Results section. I chose to present Zixuan's case because his work was typical (three out of six students did the similar thing) in the group in the sense that the production of narratives was corresponding to the task requirement. He was also very vocal in describing his work to peers during their small group discussions thus I had more data about his work. The physical ideas Zixuan brought up about electric charges and interactions were rather stable regarding the amount and kinds of ideas. Mengdi, however, was the only student in this small group who drew visual narratives throughout her worksheet whether the task explicitly asked her to draw comic strips or not. I chose to present her work because the fact that she produced narratives throughout the worksheet thus seemed to indicate that she relied on narratives a lot to describe electric charges and interactions, which is relevant to the research focus of this paper. I will present the narratives in her work for all the sections. Her ideas of electric charges expressed in the narratives increased from earlier to later sections.

## Results

In this section I will present Zixuan's and Mengdi's cases. When presenting each case, I begin with an overview of the student work on all six sections. Following that, I present each student's data section by section. In each section I will show student drawings of the comic strips (their visual narratives) and the interview transcripts in which the student was describing their work to me (their oral narratives). I also include clips of relevant classroom discussions if they help to clarify the student's narratives and/or ideas. Based on the data, I outline the narratives, highlight the narrative elements (character, relationship, events, etc.), and present the physical ideas the student expressed.

**Zixuan's Case**

**Overview.** Figure 1 shows Zixuan's work, organized according to sections. As shown in Figure 1, Zixuan produced comic strips for Sections 1, 2-1, and 2-2, in which he was asked to draw four-frame comic strips of the charges as characters. In each of the four-frame comics for Sections 1, 2-1, and 2-2, the frame sequence is indicated by the number of dots in a corner of each frame: top left, frame 1; top right, frame 2; bottom left, frame 3; and bottom right, frame 4. For Sections 3 through 5, in which he was not asked to draw stories, he shifted to drawing diagrams. Zixuan's work on Sections 3 through 5 was not treated as narratives because his production and explanation for those sections did not include any explicit narrative elements (character, location, or event, etc.). I will now describe Zixuan's work in Sections 1, 2-1, and 2-2.

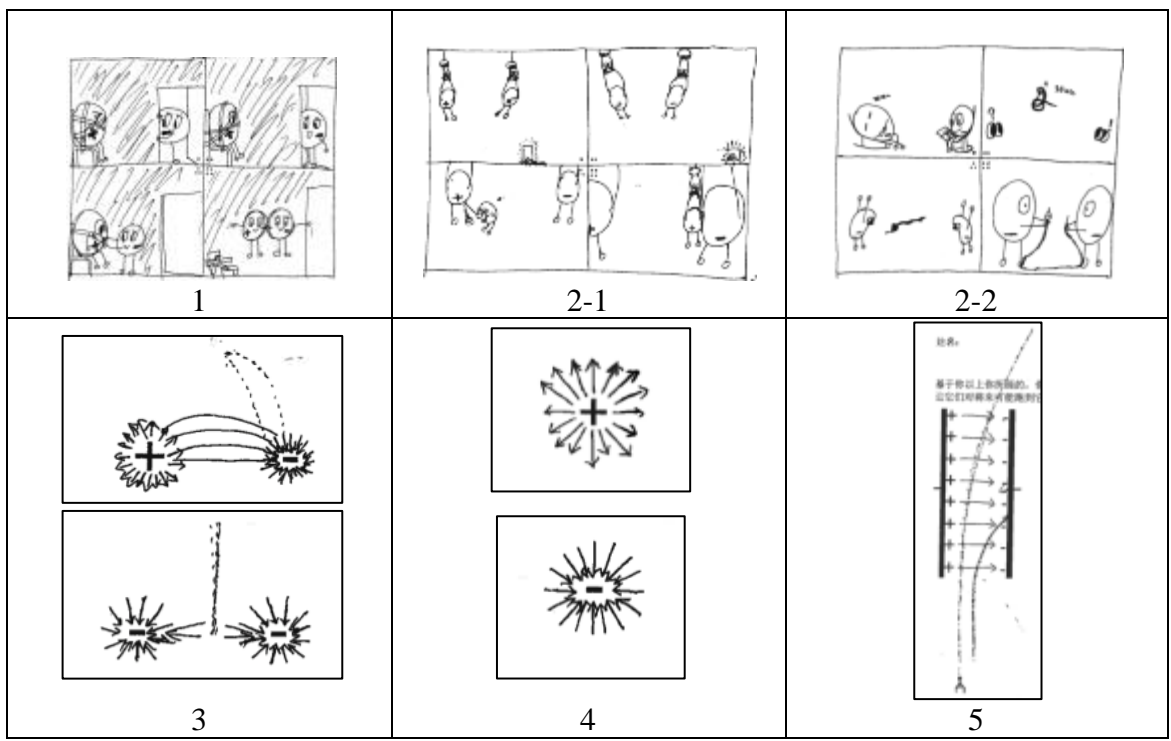

Figure 1. Zixuan's written work.

**Section 1.** Zixuan's description of his work on Section 1 is quoted below, excerpted from the interview transcript:

> This one (talking about frame 1) is just … assuming this is a room. A positive charge (the figure at the bottom left corner in frame 1) is kidnapped and strapped here, because it (the task requirement) says it (the positive charge) cannot move. Then this negative charge (the incoming figure at the top right corner) comes in through the door. When it comes in, it first holds the wall so it won't be attracted over [to the other figure]. When it completely comes in, so then the attracting force can attract them together. Nothing is in between them (referring to frame 2). This one (talking about frame 3) is that it (the incoming figure) comes here (to where the strapped figure was). The process in which they attract each other and collide with each other can be shown as the negative charge comes to rescue the positive charge. Then this one (talking about frame 4) is the two of them are like brothers and get together. Then this chair … because there is kinetic energy when it (the incoming figure) rushes in, the chair is smashed into pieces.

In this section, Zixuan constructed a narrative about one character rescuing another. He explicitly brought up some physical ideas alongside the narrative, such as: the attracting force can attract the two charges together; the positive and negative charges attract and collide with each other; and there is kinetic energy when the charge rushes in; this energy can damage things (the chair). Taking both Zixuan's written work (the drawings) and what he described for this piece of work in spoken language, the narrative, narrative elements, and the physical ideas are summarized in Table 4.

Table 4. Zixuan's Section 1.

| Narrative | Narrative elements | | Physical ideas |
|---|---|---|---|
| | Character | Two figures. A positive | |

| | | | charge and a negative charge. Each drawn as a round body with limbs. | |
|---|---|---|---|---|
| The positive charge was kidnapped and strapped in the room. A negative charge came in and rescued the positive charge. | Relationship | | Not fully specified. Mentioned near the end, "like brothers." | |
| | Time | | Not specified. Event sequence captured by the number for each frame. | |
| | Location | | A room. | |
| | Events | Initial situation | One figure was strapped in a chair. The other figure came in to rescue the strapped figure. | |
| | | Development | The incoming figure held the door. The incoming figure rushed in. The incoming figure collided with the strapped figure. The incoming figure successfully rescued the trapped figure. | The positive charge and the negative charge attract each other. The attraction makes one charge move towards the other charge and they collide. |
| | | Ending | The two figures got together. The chair was smashed down by the incoming figure. | The movement gives the charge kinetic energy. The energy can do work and damage things. |

**Section 2-1.** Zixuan described his work on Section 2-1 as follows:

This second one (Section 2-1) is…(talking about frame 1) one (the figure on the left) has positive charge and one has negative charge (the figure on the right), the two little balls (the two round-shape figures) are fixed here, because I drew them hung up. Because the two of them can produce a certain attracting force, they are a little bit toward the middle (pointing to the bottom of the two figures).

Interviewer: What are these little things (the little sticks drawn at the bottom of each figure)?

Zixuan: They're the legs.

Interviewer: Oh. Ok.

Zixuan: Then (talking about frame 2) this is a big room, and this is a door. A little positive charge comes in and sees [them], and it comes over here.

Interviewer: Over where?

Zixuan: Where the two [hanging] charges are. Then (talking about frame 3) because it… because these two (the incoming figure and the left hanging figure) are both positive charges, I thought they should know each other. When it (the incoming charge) comes to here (the left hanging charge), it (the incoming figure) gets kicked away by it (the left hanging figure), and is attracted by and stuck to the negative one (the right hanging figure), and gets hung up, too.

His description so far had not yielded a lot of explicit physical ideas, except for the idea he expressed when describing the first frame that the two hanging charges have opposite charges so they attract each other toward the middle. I followed up with questions about details in his drawings. I asked him why he drew the figures hung up.

Interviewer: What made you think of hanging them up in the first place?

Zixuan: I think…I want them (the two charges in the room) to still be able to move when they are hung up.

Interviewer: To still be able to move? (Repeats Zixuan's words.)

Zixuan: Right. Because these are one positive [charge] and one negative [charge], if I want to express that these two (the incoming figure and the left hanging figure) repel and these two (the incoming figure and the right hanging figure) attract, they (the charges) need to be able to move (to show the repulsion and attraction). So I only strapped their hands. The legs can move.

During this conversation, Zixuan expressed his idea of *two positive charges repel each other* and *one positive charge and one negative charge attract each other*. He showed this repulsion and attraction by the action of kicking away and sticking together.

Taking both Zixuan's drawings and what he described for this piece of work in spoken language, the narrative, narrative elements, and physical ideas are summarized in Table 5.

Table 5. Zixuan's Section 2-1.

| Narrative | Narrative elements | | Physical ideas |
|---|---|---|---|
| A positive charge and a negative charge were hung up to the ceiling of a room. A little positive charge came in toward the hanging positive charge, was kicked away by the hanging positive charge, and then was stuck to the hanging negative charge. | Character | Three figures: two positive charge figures and a negative charge figure, mentioned at one time as "little balls," each drawn as a round body with limbs. | |
| | Relationship | Not fully specified. Mentioned at one time that the two positive charges "know each other." | |
| | Time | Not specified. Event sequence captured by the number for each frame. | |
| | Location | A room. | |
| | Events — Initial situation | A positive charge figure and a negative charge figure were fixed, hung up, and attract each other a little bit to lean toward the middle (also toward each other). | Positive charge and negative charge attract each other toward the middle. |
| | Events — Development | A third and positive charge figure came in. The hanging positive charge figure kicked the incoming positive charge figure away. The hanging negative charge figure attracted the | Two positive charges repel each other. One positive charge and one negative charge attract each other. |

|  |  |  | incoming positive charge figure. |  |
|  |  | Ending | The incoming positive charge figure was hung up sticking to the negative charge figure. | One positive charge and one negative charge attract each other. |

**Section 2-2.** For Section 2-2, Zixuan described to me:

This one is… two negative charges (talking about frame 1). This (the figure on the left) is one of them. They (the two figures) are both fixed. This one (the left figure) is sleeping. This one (the figure on the right) is reading a book. They don't pay attention to the world around them. Then (talking about frame 2) there is a positive charge. It is a snake. It comes in, and these two (the two figures) notice him. Then (talking about frame 3) these two (the two figures) both go and pull that snake. (Interviewer: Okay… then?) Then (talking about frame 4) but none of them (the two figures) gets it (the snake).

He did not mention as many physical ideas in this section, so his ideas about electricity were not clear to me. During the interview, I probed him about his physical ideas. I asked him why he drew a snake. He answered that he wanted to draw something that could make both the positive charges nervous so they would grasp it firmly and pull it to their own sides. I followed up with why he wanted to draw that (the two positive charges being nervous and pulling), and he answered:

Because these two [figures] are both negative, and this incoming one is positive, both of the two negative ones exert attracting force on the positive one.

Interviewer: Do you mean you want to show that attraction by drawing these (the three characters in this section)?

Zixuan: Right.

During this part of the interview, he made more explicit the idea that *a positive charge and a negative charge attract* to show the two negative charges' attention to the little positive charge and the two negative charges' pulling on the positive charge. He also showed the idea of *two balanced forces cancel out in opposite directions* ("none of them get it [the snake]").

Taking both Zixuan's written work (the drawings) and what he described for this piece of work in spoken language, the narrative, narrative elements, and physical ideas are summarized in Table 6.

Table 6. Zixuan's Section 2-2.

| Narrative | Narrative elements | | Physical ideas |
|---|---|---|---|
| Two persons (negative charges) were in the room. A snake (positive charge) came in. The two persons both saw the snake. They stood up, caught the snake, and pulled it toward themselves in opposite directions. | Character | Three figures: two negative charge figures, each drawn as a round body with limbs and (in one frame) only the eyes. One positive charge figure, drawn as a snake. | |
| | Relationship | Not specified. | |
| | Time | Not specified. Event sequence captured by the number for each frame. | |
| | Location | A room. | |
| | Events — Initial situation | A negative charge figure was reading. The other negative charge figure was sleeping. | |
| | Events — Development | A third and positive charge figure came in and caught the other two figures' attention. The two negative charge characters noted the snake and both stood up. | A positive charge and a negative charge attract. |
| | Events — Ending | The two negative charges pulled the snake from the two ends. None of them got the snake. | Two balanced forces cancel out in opposite directions. |

**Wrap up.** In Zixuan's work on Sections 1, 2-1, and 2-2, he used multiple ways to express his idea of *the attraction of opposite charges*: in Section 1, this idea was represented by a negative charge coming to help a positive charge; in Section 2-1, the same idea was indicated by the two hanging charges dangling toward each other in the second scene and by the incoming positive charge sticking to the negative charge in the last scene; in section 2-2, this idea was represented by the two negative figures' attention caught by the snake (gesturing hands-up in the third frame) and by them grabbing the snake firmly (in the fourth frame). He also represented the idea of *like charges repel* in Section 2-1 by drawing one positive charge being kicked away by another positive charge.

According to the classroom videos, when Zixuan did his work, he did it independently, without talking too much to his neighbors. Usually he finished the different sections ahead of other students in his group. After he finished a section and the activity moved on to small group sharing, he showed his work to other students and described the story to them several times. The description was similar to what he described to me during the interview. His peers showed interest in his work and sometimes asked questions such as "Why did you draw them strapped?" when he described his Section 1. He referred to the task requirement that the positive charge should not be able to move to explain the reason why he drew the charge strapped to a chair. Overall, during the small group discussion times, Zixuan was presenting his work to and taking questions from the students in his group.

His ideas about electricity were relatively stable. He decided to produce visual narratives or diagrams according to the task demands. When the tasks asked him to draw comic strips, he drew visual narratives. When the tasks did not ask him to draw comics, he switched to draw diagrams. Compared to Zixuan, Mengdi's case shows very different features, which I will describe in detail in the following section.

**Mengdi's Case**

**Overview.** Mengdi's work is shown in Figure 2. For Sections 1, 2-1, and 2-2, Mengdi drew comic strips according to the task requirement. In each of the four-frame comics for Sections 1, 2-1, and 2-2, the frame sequence is indicated by the number of dots in the bottom right corner of each frame: top left, frame 1; top right, frame 2; bottom left, frame 3; and bottom right, frame 4. For Sections 3 to 5, although the tasks no longer asked her to draw comic strips, Mengdi continued to draw visual narratives and to include some narrative elements (characters, relationships, events, etc.) in her work.

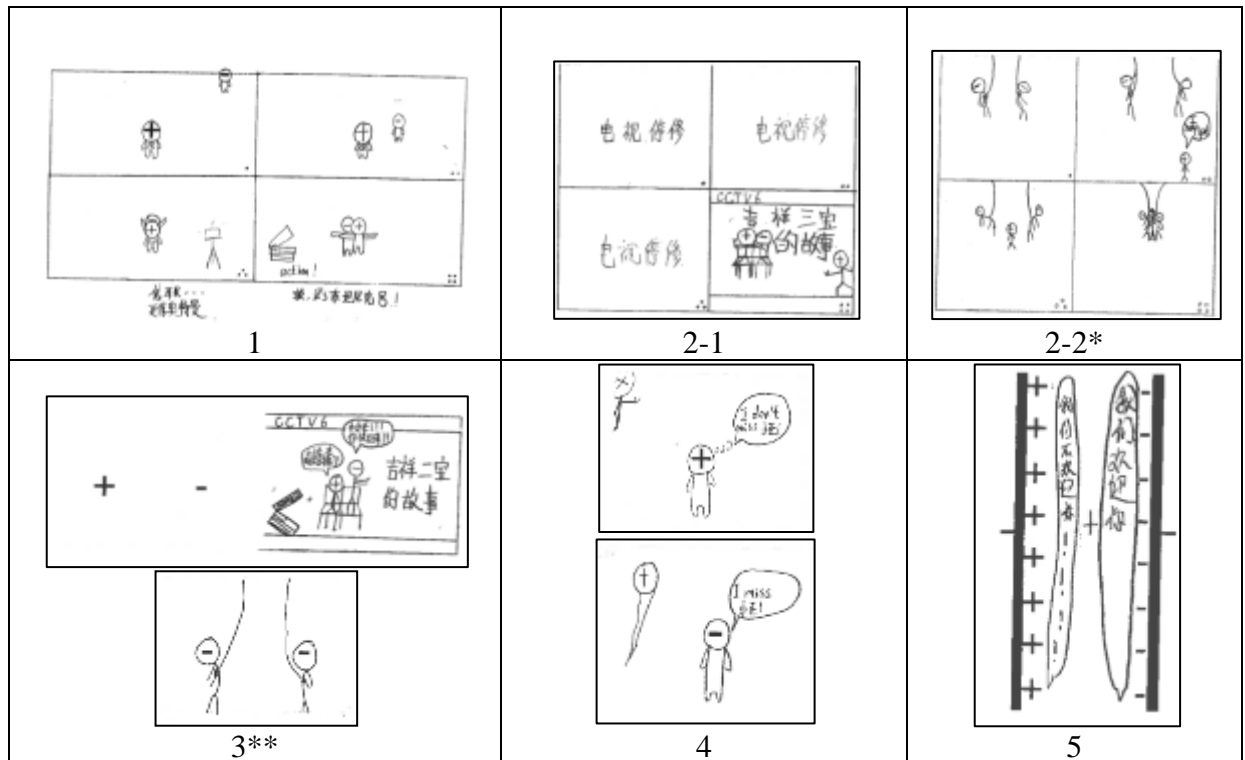

*Frames 2-4 completed during the interview. ** The top drawing was completed in class, and the bottom one was added during the interview.

Figure 2. Mengdi's written work

**Section 1.** For Section 1, Mengdi described her work to me as follows:

This is… This one [figure] is called Little Positive, and the other is called Little Negative. Little Positive stays in the room without moving (talking about frame 1). Then little Negative comes in through the door. This (frame 1) is when it (pointing at Little Negative) is at the door. This (frame 2) is when it (pointing at Little Negative) comes closer and closer to this (pointing at Little Positive). Then this (frame 3) is it (one figure) stands behind [the other figure] and gestures rabbit ears. They both... (inaudible). But this camera (in frames 3 and 4) ... is not drawn very well...

Interviewer: So here (in frame 3) you wanted to draw a camera?

Mengdi: Yeah… Like people are filming movies. This (talking about frame 4) is [that the filming] starts: "Action!" (Referring to the English word written inside frame 4). And then they are like Jack and Rose in [the movie] Titanic.

The Chinese subtitle under frame 3 reads, "We are *Ultraman*!" (Japanese television series.) The Chinese subtitle under frame 4 reads, "Oh, it is *Titanic*" (movie name).

According to the classroom video, in the beginning of the activity Mengdi encircled positive and negative signs to represent her characters. Ru, the girl sitting next to Mengdi in their small group, saw Mengdi's drawing and suggested that Mengdi should draw charges as little people. After listening to Ru's suggestion, Mengdi drew the charges as little people. When Mengdi completed her work, she did not include a camera in frames three and four. Ru looked at Mengdi's work and said the two people in Mengdi's work looked like they were posing to take pictures. Mengdi giggled and said, "Well, then I'll add a camera," which she did.

In her description for Section 1, Mengdi did not mention any electricity ideas at all. During the interview after she described the whole story to me, when I asked her if her story had anything to do with electric charges, she said that she did not think of the two figures as electric charges but just as two people. The transcript is quoted below:

> Interviewer: Well... Then does this whole story have anything to do with their positive and negative charges?
>
> Mengdi: I did not think about that then.
>
> Interviewer: Ok. So you just thought they are two persons.
>
> Mengdi: Right.
>
> Interviewer: But you did not think about their positive or negative charges.
>
> Mengdi: I didn't.
>
> Interviewer: Then what made you think that they would come closer and closer?
>
> Mengdi: Mm... Because it (one figure) came in and wanted to play with it (the other figure), they moved closer.
>
> Interviewer: Because it wanted to play with it.
>
> Mengdi: Right.
>
> Interviewer: But you did not consider that one is positive and the other is negative.
>
> Mengdi: No.

Mengdi's narrative and narrative elements for Section 1 are summarized in Table 7. She did not bring up any physical ideas regarding electric charges and interactions, even after I probed her.

Table 7. Mengdi's Section 1.

| Narrative | Narrative elements | | Physical ideas |
|---|---|---|---|
| The little positive charge was in the room. A little negative charge | Character | Little Positive and Little Negative. | |
| | Relationship | Not specified. Mentioned in the last two frames that they were characters in a TV | |

| | | came in and they took pictures together. | | series or in a Movies, e.g., Jack and Rose (a couple in a romantic relationship). | |
|---|---|---|---|---|---|
| | | | Time | Not specified. | |
| | | | Location | A room (a filming studio). | |
| | | | Events | Initial situation | Little Positive was in the room.<br>Little Negative came in through the door. | |
| | | | | Development | Little Positive came closer to Little Negative.<br>Little Positive posed together with Little Negative. | |
| | | | | Ending | The two figures were filming a movie of a romantic situation. | |

**Section 2-1.** According to the classroom video, in this section Mengdi started by constructing four very big frames that almost filled the whole page and drew something in the frames. She soon erased that and started over. Then she drew something different but soon erased that too. After several times of this back-and-forth, she erased the whole page, including the frames, and drew four smaller frames as in Figure 2, Section 2-1. In the first three frames, Mengdi drew identical pictures for each: a TV screen with the following words on the screen, "The TV is broken." In the fourth frame, she drew the TV screen, which is working. She put the label identifying the TV station in the top left corner: "CCTV 6." On the TV screen, she drew a scene of a family (a mom, a dad, and a son). The big positive and negative charges were Mom and Dad sitting on a bench. The little positive charge was their son standing in the corner. The words in the middle read, "A Story of the Lucky Three." Mengdi's explanation of her drawing was:

> This is… Back then [in class] I felt that if I filled in all the four frames with TVs, time is not enough. So I filled the first [three] ones with "the TV is broken." The last one (talking about frame 4) is a story of "The Lucky Three." They (the two figures on the bench) sit in the room, and they are fastened in their own seats. They can't move. Mm... Like... The little charge can move freely, so it makes a "V" sign [by his fingers] in front of the camera, calling for (the camera's or the audience's) attention. These two (figures on the bench) are together and cannot move. ... is what CCTV... (inaudible) They are broadcasting the story of "the Lucky Three."

During the interview, I probed about the relationship among the three figures she drew in the fourth frame:

Interviewer: What's the relationship between this little one and the other two?

Mengdi: It's their son. And he is a boy, because like things repel, and opposite things attract. This one (the mom on the bench) is his mother, because usually men are positive signs. Women are negative signs.

Interviewer: So Mom and Dad represent negative and positive (charges)?

Mengdi: Right.

Interviewer: Then this positive [charge] (the son) you think is...

Mengdi: Opposite to that (the mom), but it (the son) is smaller.

Interviewer: So you think it's the son.

Mengdi: Right.

During the above conversation with the interviewer, Mengdi mentioned her idea of *like charges repel and opposite charges attract*. She also clarified the types of charges in relation to the gender of the characters: men (the dad and the son) represent positive charges, and women (the mom) represent negative charges. It was not clear at the time how she represented the attraction and repulsion in her work. My interview questions following the above conversation prompted her to talk more about the actions and interactions between the figures:

> Interviewer: And he (son) has been standing there all the time?
>
> Mengdi: He is like… previously wanted to stand with them (the mom and dad), but he ran to the front to attract the camera's attention, so...
>
> Interviewer: Oh... I see. How did he run, just by itself? (In retrospect, this was not a well-articulated question.)
>
> Mengdi: Right. He might be... I didn't think of that. But he might be kicked over there by his dad. Or maybe [because] charges repel.
>
> Interviewer: Repelled by which one?
>
> Mengdi: By his dad's positive charge.
>
> Interviewer: What about his mom?
>
> Mengdi: Mm... His mom stays there still. Because he (the son) stands on the right, he is close to her. The strength is strong, and he pushes him to here (this sentence probably shifted to talk about the dad again, possibly referring to the dad pushing the son. This is not clear from the transcript). I did not think about that then. I just wanted to show he (the son) wanted to be on camera.

Her explanation about how her ideas of charges' attraction and repulsion relate to the narrative was still not clear to me, but I did not want to push her too hard so I moved on to the next section. Nonetheless, she explicitly said that *like charges repel and opposite charges attract* in relation to the characters for a couple of times. She did not do this in her description for Section 1. Her narrative, narrative elements, and physical ideas are summarized in Table 8.

Table 8. Mengdi's Section 2-1.

| Narrative | Narrative elements | | Physical ideas |
|---|---|---|---|
| A family of three (mom, dad, and the son) was filming a show. | Character | Two big charges and one little charge. | |
| | Relationship | Mom, dad, and son. | Male represents positive charge. Female represents negative charge. |
| | Time | Not specified. | |
| | Location | They were filmed and shown on a TV screen. | |
| | Events | Initial situation | Not shown. | |
| | | Development | Now shown. | |
| | | Ending | The three figures as a family were filming a show in front of a camera. The mom and dad were sitting together on a bench. The son was standing at the bottom right corner, at the front. The son was closer to his mom. The son was kicked away by his dad. | Like things repel. Opposite things attract. |

**Section 2-2.** She did not get to complete Section 2-2 in class, but had drawn the first frame as shown in Table 7. During the interview, I asked her to finish it. She described the narrative to me before she drew:

Mengdi: I didn't finish this (Section 2-2) in class. This (frame 1) is just scribbles...

Interviewer: No worries. You can tell me how you would have drawn it (hands her a pen) if you had the time.

Mengdi: [This is] a story about two big charges and one small charge. These two [big charges] are hung up (to the ceiling). Then a little positive charge comes in. It wants to help both [of the two big charges]. Because opposite things attract, the three of them attract together. Because there is the little charge in the middle, the three of them are together. The ending is similar to one of the previous ones (we had just looked at some other students' work for Section 2-1 before we started to talk about her own work for Section 2-2). Similar to Zixuan's work in which the little positive charge was kicked away and none of the three got rescued (referring to Zixuan's work for Section 2-1), this one (referring to her own work for Section 2-2) is that they all attract together. The two ones (the two hanging charges) repel each other so they are a little bit outwards, previously (before the little positive charge comes in). In the end the little positive charge attracts the other two toward the middle. Because the little positive charge is closer (to the two charges), the little positive charge can't come down although he wants to (the little positive charge was clipped between the two big negative charges).

After she described her thinking, she drew the three frames to complete the comic strip, as shown in Figure 2, Section 2-2.

In her description, Mengdi mentioned the repulsion between the two hanging negative charges: *The two ones (the two hanging charges) repel each other so they are a little bit outwards, previously.* She also mentioned the attraction between both of the hanging negative charges to the incoming positive charge: *Because opposite things attract, the three of them attract together; the little positive charge attracts the other two toward the middle.* She got two ideas about the hanging charges: (1) there are two negative charges, so they repel each other and separate apart; (2) they are negative charges, so they are attracted to a positive charge in the middle and squeeze toward the middle. At the end of the above excerpt, she was trying to coordinate these two results by addressing the relationship between charges' distance and forces' strength: *Because the little positive charge is closer (to the two negative charges), the little positive charge can't come down although he wants [to].* Although this issue was not totally resolved yet in her description, she expressed the idea that the attraction between a hanging, negative charge and a incoming, positive charge is stronger than the repulsion between the two hanging, negative charges, because the two attracting charges are closer in distance. With some probing, she talked more about the physics underlying these events:

> Interviewer: Why did you originally draw them hanging up (frame 1)?
>
> Mengdi: Because like... I don't know... I feel like... Drawing a chair (this was probably what she had thought of drawing but ended up not doing this) is a little too difficult. So I hung them up. This way it is easier to show that if they are hung up in the air, they repel, they repel more (gestures a separation of two hands)...

Mengdi seemed to be indicating that she tried to find a way to show the repulsion between the two hanging negative charges and ended up by drawing the two charges hanging so that they could dangle. A little bit afterwards, I asked her:

> Interviewer: What about these being closer and closer (frames 3 and 4)*?*
>
> Mengdi: They are closer because the little charge is in the middle like… (claps her hands), because the two of them (the two handing charges)... Opposite things attract, so both of them are attracted here (to the middle). And this one (frame 4) because there is a strong attracting force, they are squeezed in the middle. As a result, all three of them are hanging there. No one can come down.
>
> Interviewer: You think the attracting force is very strong?
>
> Mengdi: Yes. Because they are two big charges and a little charge.

Mengdi continued to talk about the attraction between the hanging negative charges and the incoming positive charge, and continued to say that the attracting force was strong. When I probed her about the strength of the force, she did not emphasize the distance but rather the size and number of the charges ("because they are two big charges…").

Her ideas about the relationship between the force strength, the distance, and the number/size of the charges were not completely articulated as in the canonical Coulomb's Law. However, these factors she was attending to during the interview are essential to what she was going to learn about Coulomb's Law later in high school.

Mengdi's narrative, narrative elements, and physical ideas exhibited in Section 2-2 are summarized in Table 10.

Table 10. Mengdi's Section 2-2.

| Narrative | Narrative elements | | Physical ideas |
|---|---|---|---|
| Two big negative charges were hung up to the ceiling of the room. A little positive charge came in and was clapped in the middle of the two negative charges. | Character | Two big charges and one little charge. | |
| | Relationship | Not specified. | |
| | Time | Not specified. | |
| | Location | A room. | |
| | Events — Initial situation | The two negative charges were hanging up and separate outward. | Like charges repel. |
| | Events — Development | The little positive charge came in. | |
| | Events — Ending | The little positive was clapped in the middle of the two negative charges. | Opposite charges attract. A closer distance may bring a stronger force. Charge number/size matters for force strength. |

**Section 3.** Mengdi's work for Section 2-2 was completed during the interview. In class, she continued with Section 3 after she finished the first frame of Section 2-2. Her narratives in Section 2-1 and Section 3 were similar and related (e.g., similar stories about mom, dad, and son), as shown in her drawings and explanations for both. Starting from Section 3, the task no longer asks student to draw comic strips but asks students to summarize the impacts between electric charges (in different situations) in individual frames. As shown in the first case, Zixuan stopped drawing comics and started to draw diagrams for Sections 3-5. Mengdi, however, continued to draw one-frame comics for these sections, showing a static scene of a narrative.

For Section 3, Mengdi drew a TV screen (putting TV station labels at the top) airing a show. Only the mom (represented by the negative charge) and dad (represented by the positive charge) of the family were in the scene. The mom reached her hands out to the out-of-the-picture son, "Little Positive, come back here soon." Mengdi used this to express attraction between the two charges. The dad was drawn sitting upright with his back toward the right side (where the son was supposed to be). His hands were at his waist. All these gestures indicated the anger of the dad and the repulsion between the two positive charges (dad and son). The subtitle says, "Little Positive, if you don't come back, the show can never be made!" During the interview, she explained to me:

> This one, the little positive charge, is taken away. If Little Positive is taken away, one can only make a story of two [characters] (speaking in comparison to the Story of Lucky Three that she produced in Section 2-1). And the two of them (mom and dad) miss him (the son) a lot. They are the parents. Then the mom is like "Little p, come back soon." "If you don't, the show can't move on."

In this scene, the character and relationship was similar to what was in her work for section 2-1, i.e., mom, dad, and son. The mom was still kind to the son and the dad was mad at the son. In order to represent these characters and relationships, Mengdi included multiple representations in her drawing words (e.g. mom's kind or dad's angry words to their son), characters' gestures (the mom reaching hands, the dad sitting upright and backward to the son), characters' tones (indicated by the double exclamation marks), etc. to express her ideas of *like charges repel and opposite charges attract*. She also completed Section 3-2 during the interview and explained to me:

> Then the two of them (the two negative charges)... They separate again. Because the middle one is... The two of them are like "Tseng!" (gestures two hands separate apart) to the sides.

By drawing this work, she showed the idea of like charges repelling. Mengdi's narrative scenes and physical ideas work for Section 3 and are summarized in Table 11.

Table 11. Mengdi's Section 3.

| Narrative scenes | Physical ideas |
| --- | --- |
| 3-1 Mom was nice to the absent son. Dad was mad at the absent son. | Like charges repel. Opposite charges attract. |
| 3-2. Two hanging figures separate apart. | Like charges repel. |

**Section 4.** In her drawing for Section 4, Mengdi added the little positive charge back in the scenes. In the top picture, the little positive charge stands in the corner of each picture and gestures defiantly, shaping the fingers of his left hand into a "V" and resting his right hand on his hip. The big positive charge says he does not miss the little positive charge, "I don't miss Little Positive." The words "don't miss," the defiant gestures, and the distance all represent repulsion. In the bottom picture, the negative charge says, "I miss Little Positive." The little positive charge is drawn closer to the negative charge with a tail swinging to the left, indicating the direction of moving toward the negative charge on the right. The word "miss," the direction of the swinging tail, and the distance all represent attraction. She explained her work to me:

I drew this one (the bottom picture) first. This is an imagined picture (not clear what she meant; she could mean that she imagined a situation for the characters). Because she (the negative charge) is a negative charge, and she misses little positive and wants to play with him. But this one (the positive charge), because they are the same gender, so he (the positive charge) doesn't want little positive charge. This is… like… he (the little positive charge) comes out and is mocking him with a hand gesture, like… (mimics the character and makes a "V" sign with her fingers).

In these drawings and descriptions for Sections 3 and 4, Mengdi continued to show that *like charges repel and opposite charges attract*, and was more fluent in creating situations of narratives, or scenes, to express her ideas. Mengdi's narrative scenes and physical ideas in Section 4 are summarized in Table 12.

Table 12. Mengdi's Section 4.

| Narrative scenes | Physical ideas |
| --- | --- |
| 4-1. One figure does not miss another figure. The other figure mocks it. | Like charges repel. |
| 4-2. One figure wants to play with the other figure and misses it. | Opposite charges attract. |

**Section 5.** In Section 5, Mengdi added a positive sign in the middle of the two charged boards. Next to the line of positive charges on the left, she wrote, "We don't welcome you!!!!" which was meant to represent their *repulsion* toward the positive charge in the middle. The line of negative charges on the right say to the positive charge, "We welcome you," which was meant to represent *attraction*. In addition, Mengdi circled the lines of words next to all the charges on one side, and explained that they say the words in the same voice, which could represent a rudimentary idea of superposition. She added multiple exclamation marks at the end of the words (by the left hand side board), which could mean that she tried to express that the collective repulsion was very strong.

> This one (the positive charge she added in between the boards) is, I think, a little positive charge, because here (the left hand side board) is all like charges repelling, so they all don't welcome the newly coming little positive charge. But this one (the right hand side board), negative charges, they hope little positive charge to come very much. So they are saying, "we welcome you."

Mengdi's narrative scenes and physical ideas in Section 5 are summarized in Table 13.

Table 13. Mengdi's Section 5.

| Narrative scenes | Physical ideas |
| --- | --- |
| A group of positive charges do not welcome a positive charge. A group of negative charges welcome a positive charge. | Like charges repel. Opposite charges attract. Electric force superposition. |

**Wrap up.** Mengdi's case demonstrated the affordance of narratives to describe scientific phenomena. She started with almost no (or at least no explicit) physical ideas in Section 1 and ended up with rich ideas in the end (such as like charges repel; opposite charges attract; the larger the charge amount, the stronger the force; the closer the distance, the stronger the electric force; etc.). At first, she struggled at producing narratives for the charges, and later ended up making various narratives for the ideas she wanted to express in her work. Throughout the sections, Mengdi was able to describe more ideas about electric interactions, and the ideas were more explicit. Her ideas of the electric field first focus, implicitly, on motion (two charges moving closer), then on interaction (electric attraction and repulsion), then on the more detailed factors of the electric attraction and repulsion (distance, amount of charges, force strength), and then on superposition (charges acting forces together). She was first able to represent the phenomena (positions of charges) and then also was able to represent the mechanics that caused the phenomena (attraction and repulsion that determined the change of positions). She used multiple representations in later sections: figures' positions, gestures, words, and exclamation marks. The ideas expressed—*attraction, repulsion, superposition*, etc.—were rich and becoming more and more explicit in her work.

Also, the connection between her narratives and her ideas of the electric field were more explicit in later sections. When she described her work of Section 1 during the interview, she did not mention any ideas about electric charges. Even after I explicitly asked her how the story she had constructed connected to electric charges, she told me that she did not think of the characters as charges when she was drawing the story. In contrast, in the later sections, when I asked her about the stories she constructed, she frequently referred to the physical ideas (e.g., opposite charges attract; like charges repel) as the reason why she constructed the stories as they were (e.g., Mom likes the son; Dad was mad at the son).

She completed her work more independently (no longer needing peer's facilitation) and more smoothly. In Section 1, Mengdi drew charges as people only after she took Ru's suggestion, and she used this thereafter throughout all the sections. When she was working on Section 2-1 in class, she spent a while without drawing anything down on paper. She listened to her small group as they shared their work and then drew the charges as a family. In later sections, however, Mengdi completed her work on her own and described her ideas to me with confidence during the interview (e.g., the narrative about the two figures either missing or not missing the third figure to represent that like charges repel and opposite charges attract; the narrative that charges collectively welcome or do not welcome the test charges to show the superposition of electric attraction and repulsion).

## Discussion

This study explored the narratives high school students constructed for electrostatic phenomena and the physical ideas the students expressed in their narratives. The research questions ask: given the prompt to produce narratives for electrostatic phenomena during a classroom activity prior to receiving formal instruction,

(1) what ideas of electrostatics do students attend to in their narratives?

(2) what role do students' narratives play in their understanding of electrostatics?

In the Results section I presented two cases (Zixuan's case and Mengdi's case) with details regarding their narratives and ideas to answer my research questions and to support my main argument: oral and visual narratives can become tools for high school students to make sense of concepts such as the electric field. Below I will summarize my findings and draw educational implications.

**The Physical Ideas**

By creating narratives, the high school students in this study expressed ideas of the electric field that included: the rule that opposite charges attract and like charges repel (in both students' multiple sections); the strength of the electric attraction and repulsion in relation to the charge amount and distance (Mengdi in Section 2-2); the speed and direction of a moving charge in presence of other charges (Zixuan in Sections 1 and 2-1); the simultaneity of interaction among multiple charges (Mengdi in Sections 2-1 and 2-2, Zixuan in Section 2-2); the balance between two competing and balanced forces produced by the opposite charges (Mengdi in Section 2-2, Zixuan in Section 2-2); the superposition of multiple forces (Mengdi in Section 5).

**Narratives as A Tool to Express Ideas**

Previous studies have indicated that narratives can be a helpful tool for students to express their ideas of scientific phenomena (e.g., Sohmer and Michaels, 2005). In Sohmer and Michaels's work (2005), a specific narrative tool, the Little Puppies story, was introduced by the teacher and was taught to middle school students. Students in Sohmer and Michaels' study learned to use the Little Puppies tool to make sense of the behavior of air molecules and the kinetic theories of gases. The designer of the Little Puppies narrative tool made the connection between the characteristics of air molecules and the characteristics of the animated character, Little Puppies. The mindless agents (little puppies) represented the air molecules. They were mindlessly moving around and their behaviors (position and velocity) were determined by the external condition (temperature, pressure, and/or size of the container, etc.). The students in Sohmer and Michaels' study learned to use the tool to make sense of the behavior of air molecules (2005).

In the comic strip activity featured in the present study, students did most of the work making connections between the characteristics of electric charges and the characteristics of the cartoon characters they created in their individual work. The comic strip task set up a beginning scenario about electric charges and asked students to extend the story and to fill in multiple comic panels. In the task description of the comic strip activity, the electric charges were named Little Positive, Little Negative, Big Positive, and Big Negative. The source charges were described as staying in a room and immobile. The test charges were described as coming in the room where the source charges(s) rested. The beginning scenario was given to the students and it set up students to describe, in a storyline, the behavior of the test charge according to the interaction between the test charge and the source charge(s).

The comic strip task created a space for students to come up with ideas of social relationship, emotion, attitude, and/or action to connect to characteristics of electric charges (such as attraction, repulsion, force strength, and force superposition). Compared to the Little Puppies narrative tool in Sohmer and Michaels's work (2005), the narrative tools students created in the comic strip activity were more related to each student's individual experience and interest. The narratives students constructed served as a tool for them to describe, make sense of, and communicate ideas of the physical concepts in question. The narratives also provided researchers a venue to learn about the students' ideas.

**The idea of being static.** In electrostatics the electric field constructed by source charges is "static," meaning the electric field does not change over time. In the situation described in the comic strip worksheet the electric field was static because source charges' positions were fixed—the source charges were staying in the room and unable to move. Students, therefore, drew upon their social ideas about characters to address the immobility of these charges.

In his work of Section 1 Zixuan created a situation of kidnapping and a scene of the character strapped in the chair. In his work of Section 2-1 he created a situation that the characters were hung up to the ceiling. We see that in both cases Zixuan's idea of immobility was associated with his idea of lack of freedom. In Section 2-2, Zixuan created a scene of two people reading and sleeping, in which his idea of immobility was associated with the idea of being focused, quite, and unaware of the outside world. Mengdi, in her work, created scenes of photo taking and/or film shooting, in which the characters were asked to pose in front of a camera thus they stayed still for a short period of time. Mengdi's work showed that her idea of immobility was associated with the idea of posing in front of a camera.

**The idea of attraction and repulsion.** When two electric charges interact, like charges repel and opposite charges attract. Each charge experiences a force, repulsion or attraction, from the other charge. The electric forces are invisible. In their work students addressed these electric forces and made the attraction/repulsion visible in the narratives.

In his work Zixuan described the electric attraction and repulsion in multiple ways. In Sections 1, the second frame, Zixuan drew the incoming negative charge firmly holding the door frame (hands on the frame, exaggerated facial expression such as widened eyes) to avoid being pulled over to the positive charge in the room. The attracting force between the two charges was invisible, but its counter part—the door frame holding—was drawn visible in the characters' gesture and facial expression. In the following frame (the third frame) Zixuan drew that the incoming negative charge released the door frame and rushed rapidly toward the positive charge in the room (smashed the chair when arriving at the positive charge). In this third frame, Zixuan's idea of electrically attraction was represented by its effect: the visible rapid motion. In his work of Section 2-1 Zixuan drew a movable leg of the positive source charge so that it was clearer to the reader who exerted the repelling force on the test charge. In his work of Section 2-2 Zixuan represented the electric attraction in two ways: the characters' emotion (being scared and nervous upon seeing the snake) and their action (pulling the snake in the middle).

In most parts of her work Mengdi represented electric interactions through human emotion, attitude, and kinship. She represented the idea of electric attraction by describing a mom liking the son, being kind to the son, and missing the son when he was absent. She also represented the idea of electric repulsion by describing a dad being mad at the son and not missing the son when he was absent. Mengdi drew out these emotions and attitudes in the characters' gestures (e.g., mom's hands out to illustrate the missing, dad's sitting back toward the son to illustrate being mad) and in bubbles of characters' words (e.g., the character says "I miss you"). In Section 5 of her work Mengdi described the attraction as welcoming and repulsion as not welcoming. She added word bubbles (e.g., "We welcome you.") to describe these situations and sometimes included punctuations to highlight the strength of the attitude (e.g., repeated exclamation marks in "We don't welcome you!!!!").

**The idea of force superposition.** When multiple source charges simultaneously exert forces on a test charge, the effect of all the electric forces can be described by vector superposition, adding up all the force vectors and getting an equivalent net force. As a result the test charge moves as if it experiences one force, the net force. None of the students participated in this study had officially learned vector superposition at the time of the study, yet in their work the students started to express some ideas about the qualitative aspects of force superposition, such as multiple forces act together; all the forces collectively determine the effect on a test charge. In some parts of their work, some of the students also started to express the quantitative aspects of force superposition, such as the net force of two equal and opposite forces is zero.

In her work of Section 2-1 Mengdi described that the mom and dad each had their respective attitude to their son (the attracting and repelling forces from different source charges on the test charge). In her work of Section 5, Mengdi drew multiple charges speaking in groups that they welcome (attraction) or not welcome (repulsion) the test charge. These drawings exhibited some understanding of multiple forces acting forces together on the test charge. The effect of these forces on the test charge was not addressed clearly in Mengdi's work.

In his work of Section 2-1, Zixuan drew that the positive source charge kicked the test charge away (repulsion) while the negative source charge glued the test charge (attraction). The test charge, therefore, ended up sticking tightly with the negative charge and was forced to also be hung up to the ceiling. In this piece of work, both ideas were exhibited: multiple forces act together and all the forces collectively determine the effect on a test charge. In his work of Section 2-2, Zixuan drew that the two persons were pulling the snake in opposite directions and the snake ended up not moving anywhere but staying in the middle. The piece of work showed some understanding that when the two forces were equal and opposite, the net force was zero and the test charge would not change its position.

In sum, the data made us able to see how students' ideas of physical concepts were deeply associated with their ideas of everyday events. The opportunity for students to create their own narratives for the physical scenario in question allowed them to connect the physical concepts with their personal experience and interest, and allow us as researchers to recognize different students' thinking in the same context.

**Visual and Oral Narratives**

Previous studies of science education about narratives have been focused on spoken narratives (Ochs, 1992; Sohmer & Michaels, 2005) or written narrative text (Norris, 2004). In many situations people express more ideas if they were given the opportunity to also produce visual narratives (Keats, 2009). Keats developed a methodology of narrative analysis that includes spoken, written, and visual narrative texts (2009). The methodology has been applied to studying narratives of personal experiences, but not yet to studying narratives in a science education context. The present study connected these areas and included both visual and oral narratives to study students' understandings of electrostatic interactions in a classroom context. The comic strip drawing task prompted students to produce visual narratives on the worksheet. The group discussion during the activity and the post-class interview encouraged students to produce oral narratives. Including multiple formats of narrative students can employ more modalities to express their ideas. Researchers, therefore, have more venues to learn about students' ideas.

In his work of Sections 1 and 2-1 Zixuan drew ropes strapping the figures when he described that big charges' positions were fixed. He added little legs to the figures to show that the source change can exert repelling forces. In his work of Section 2-2, in order to express his idea of electrostatic attraction, Zixuan drew the facial expressions and gestures of the figures (big eyes, raising arms) to describe that the charges were scared and their attention was caught by the incoming charge (the snake). Producing the drawings Zixuan was able to put his ideas on paper. On the other hand, his oral explanation of his work (during the activity as well as during the interview) can clarify his ideas and his drawings when the reader cannot make sense of his work by solely reading his work.

Mengdi drew the mom reaching her hands out to the son when describing attraction. She also drew the dad sitting with his back toward the son when describing repulsion. Comic strips allowed her to add words ("miss," "not miss," "welcome," "not welcome") and symbols (e.g., punctuations) to make it more explicit and/or more conspicuous her ideas in the work.

**Implications for Education**

The present study implies that narrative activities can be more productive for high school students' learning of science than we might have previously considered. The data of the present study showed that students' ideas of physical concepts (such as electrostatic interaction) were deeply associated with many other ideas, such as their social ideas of human interaction. Curriculum writers can develop materials addressing these connections so that the learning of physical concepts does not appear to be an isolated conception with no connection with a learner's personal experience and perspective.

Educator can design and run classroom activities encouraging students to articulate their understandings of physical concepts in narrative languages besides in formal disciplinary languages. This way the activity can be oriented to productively sense-making rather than only reciting procedures of problem solving.

In a classroom the teacher can attend to these narrative ideas when students bring them up. The teacher can initiate productive questions and/or discussions based on these narrative ideas, and then develop further discussion topics guiding students to think deeper about the scientific concepts in question.